\documentclass[aps,preprint,showpacs,superscriptaddress,groupedaddress]{revtex4}  
\usepackage{graphicx}  
\usepackage{dcolumn}   
\usepackage{bm}        
\usepackage{amsmath,amssymb}   

\usepackage{epstopdf}
\usepackage{hyperref}
\hypersetup{colorlinks=true, linkcolor=blue, urlcolor=blue, citecolor=red, bookmarks=false, pdfstartview={FitH}}

\begin{document}

\widetext

\title{Quasibound states of the Dirac field in Schwarzschild and Reissner-Nordstr\"om black hole backgrounds}

\author{Ciprian A. Sporea}%
 \email{ciprian.sporea@e-uvt.ro}
\affiliation{\small \it
 West University of Timi\c soara, V.  P\^ arvan Ave.  4, RO-300223 Timi\c soara, Romania}

\begin{abstract}
In this paper we study the existence of (quasi)bound states in two spacetime geometries describing Schwarzschild and Reissner-Nordstr\"om black holes. For obtaining this type of states we search for discrete quantum modes of the massive Dirac equation in the two geometries. After imposing the quantization condition, an analytical expression for the energy of the ground states is derived. The energy of higher states is then obtained numerically. For very small values of the black hole mass $M$ we compare the energy of the Reissner-Nordstr\"om black hole quasibound state with the Dirac-Coulomb energy and we have found the two to be in good agreement.

\end{abstract}

\keywords{quasibound states; massive Dirac field; black holes.}

\pacs{04.70.-s, 04.62.+v.}
\maketitle

\section{Introduction}

The propagation of quantum fields with zero or non-zero spins in different black hole backgrounds were intensively studied in the literature. The vast majority of these studies made use of numerical techniques to investigate the specific task at hand. Moreover, due to the complexity of the equations involved, analytical methods were used less often and only for certain problems. For example, although the problem of fields scattering by different types of black holes received much attention \cite{futterman,dolan,dolan11,brito,unruh,Rogatko,Gaina1,Matzner1,Matzner2,Ford1,Ford2,konoplya,sanchez,sanchez1,anderson,batic,crispino11,crispino12,oliverira}, only recently the authors of Refs. \cite{sporea1,sporea2,sporea3,wu-dai1,wu-dai2} have found analytical phase shifts.

The configurations that a quantum field could take in the surroundings of a black hole depend on the specific boundary conditions imposed to the field at the black hole horizon, respectively at infinity. All these configurations can have complex frequencies who's real part (magnitude) give rise to oscillations rates and the imaginary part is related to the rate of decay of the states. Thus one can have quasinormal modes that are purely ingoing at the horizon and outgoing at infinity. These are the most studied configurations \cite{QNM1,QNM2,QNM3,QNM4,QNM5,QNM6,QNM7,QNM8,QNM9,QNM10,QNM11,QNM12,QNM13,QNM14,QNM15,QNM16,QNM17,QNM17a,QNM18,QNM19,QNM20}. One can also have stationary resonances and dynamical resonances (for a detailed discussion on these configurations see Refs. \cite{barranco, herdeiro} and also Ref. \cite{xiang-nan}). This problem is also discussed in Refs. \cite{rez1,rez2,rez3,rez4,rez5,rez6}. The existence of pure gravitational bound states (the analogue of the electromagnetic bound states) is believed do not exist in black hole geometries due to the singularity present, that acts as a current sink. However, quasibound states that can be long lived do exist and for the Dirac field refs. \cite{dolan1,gaina1,gaina2,gaina3} were among the first papers that studied them. Other studies of quasibound states in different black hole spacetimes can be found in refs. \cite{cardoso,dolan2,dolan3,dolan4,Ruffini,crispino,crispino2a,bond1a,bond2a,bond3a,bond4a,bond5a,bond6a,bond7a,bond8a,bond9a,bond10a}.

In this paper we will study the quasibound states of the massive Dirac field in Schwarzschild and Reissner-Nordstr\"om black holes, with a focus on the last one. Using the discrete spectrum of the Dirac equation in these black holes we derive analytically the energy of the quasibound states. The formula obtained for the energy is only an approximative one so that we are restricted to make only a qualitative analysis of the quasibound state spectrum. The organization of the paper is as
 follows. In Sec. \ref{sec2} we discuss the Dirac equation in a Reissner-Nordstr\"om background and the derivation of the discrete quantum modes, while in Sec. \ref{sec3} using a quantization condition an approximative formula for the energy of the quasibound states in this geometry is obtained. In the next Sec. \ref{sec4} we discuss the results obtained and the final Sec. \ref{concl} is kept for Conclusions.

\section{Solutions of the Dirac equation: discrete quantum modes}\label{sec2}

The first solution of an electrically charged black hole was the Reissner-Nordstr\"om solution, having as the line element the following expression
\begin{equation}\label{ec1}
ds^2=h(r)dt^2-\frac{dr^2}{h(r)}-r^2\left( d\theta^2+\sin^2\theta d\phi^2 \right)
\end{equation}
with
\begin{equation}\label{ec2}
h(r)=1-\frac{2M}{r}+\frac{Q^2}{r^2}
\end{equation}
and were $M$ stands for the mass of the black hole and $Q$ it's total electrical charge. By setting $h(r)=0$ we find that the RN black hole has two horizons, a Cauchy inner horizon ($r_-$) and an outside event horizon ($r_+$), given by
\begin{equation}\label{ec3}
r_{\pm}=M\pm\sqrt{M^2-Q^2}
\end{equation}
Being electrically charged the black hole will produce in it's exterior an electrostatic potential of Coulomb type $Q/r$ and a fermion with elementary electric charge $e=\pm\sqrt{\alpha}$ \footnote{In this system $\alpha\simeq \frac{1}{137}$ is the fine structure constant while  the electron mass is $m_e=\sqrt{\alpha_G}\simeq 4.178\, 10^{-23}$.} will have the following potential energy
\begin{equation}\label{ec4}
V(r)=\frac{eQ}{r}
\end{equation}

Using the Cartesian gauge \cite{Vilalba,cota1} it was showed that in a curved space-time of a spherically symmetric black hole \cite{cota1,sporea1,sporea2} the Dirac equation $i\gamma^{a}D_{a}\psi - m\psi=0$ can be separated into an angular and radial part. The angular part turns out to be the same as in relativistic flat spacetime theory, thus that the fundamental solutions for the particle-like energy eigenspinors of energy $E$ will be of the form
\begin{equation}\label{ec5}
	\begin{split}
	\psi(x)=&\psi_{E,j,m,\kappa}(t,r,\theta,\phi) =\\
	&\frac{e^{-iEt}}{r\,h(r)^{1/4}}\left\{ f^+_{E,\kappa}(r)\Phi^+_{m,\kappa}(\theta, \phi) + f^-_{E,\kappa}(r)\Phi^-_{m,\kappa}(\theta, \phi) \right\}
	\end{split}
	\end{equation}
were $f^\pm_{E,\kappa}(r)$ are the unknown radial wave functions, while $\Phi^\pm_{m,\kappa}(\theta, \phi)$ are the usual 4-component angular spinors \cite{Greiner, Landau4} and for $\kappa$ we use the convention $\kappa=\pm(j+1/2)$ with $l=|\kappa|-(1-sign\,\kappa)/2$.

The reaming radial part of the Dirac equation can be brought to a Hamiltonian form $H_r{\cal F}=E{\cal F}$, namely
\begin{equation}\label{ec6}
\renewcommand{\arraystretch}{1.8}
\left(\begin{array}{cc}
    m\sqrt{h(r)}+\frac{\textstyle eQ}{\textstyle r}& -h(r)\frac{\textstyle d}{\textstyle dr}+\frac{\textstyle \kappa}{\textstyle r}\sqrt{h(r)}\\
h(r)\frac{\textstyle d}{\textstyle dr}+\frac{\textstyle \kappa}{\textstyle r}\sqrt{h(r)}& -m\sqrt{h(r)}+\frac{\textstyle eQ}{\textstyle r}
\end{array}\right)
\left(\begin{array}{c}
f^+_{E,\kappa}(r)
   \\
f^-_{E,\kappa}(r)
\end{array}\right)
=E\left(\begin{array}{c}
f^+_{E,\kappa}(r)
   \\
f^-_{E,\kappa}(r)
\end{array}\right)
\end{equation}

The above exact radial problem has no known analytically solutions. For obtaining analytical solutions we are forced to resort to a method of approximation. We will focus here on finding the discrete quantum modes that allow for the existence of quasibond states in a region far away from the Reissner-Nordstr\"om black hole. The continuous modes and the scattering of fermions by Reissner-Nordstr\"om black holes were discussed in our previous paper \cite{sporea2}.

After introducing the Novikov variable
\begin{equation}\label{ec7}
x=\sqrt{\frac{r}{r_{+}}-1}\,\in\,(0,\infty)
\end{equation}
one multiplies the resulting equation with $x^{-1}(1+x^2)$ and then for large values of $x$ take a Taylor expansion with respect to $\frac{1}{x}$ of the resulting equation. Keeping only the dominant terms and neglecting the terms of the order $O(1/x^2)$ and higher, the radial Dirac equation (\ref{ec6}) will reduce to \cite{sporea2}
\begin{equation}\label{asimpt2}
\renewcommand{\arraystretch}{1.8}
\left(\begin{array}{cc}
\frac{\textstyle 1}{\textstyle 2} \frac{\textstyle d}{\textstyle dx}+\frac{\textstyle\kappa}{\textstyle x}
& -x(\mu+\varepsilon)- \frac{\textstyle 1}{\textstyle x}(\zeta+\beta)\\
x(\varepsilon -\mu)-\frac{\textstyle 1}{\textstyle x}(\zeta-\beta)
& \frac{\textstyle 1}{\textstyle 2} \frac{\textstyle d}{\textstyle dx}-\frac{\textstyle \kappa}{\textstyle x}
\end{array}\right)\left(\begin{array}{c}
f^+(x) \\
f^-(x)
\end{array}\right)=0
\end{equation}
where we have introduced the notations
\begin{equation}\label{asimpt3}
\begin{split}
&\zeta=\frac{m}{2}(r_+-r_-)=\frac{1}{2}\,\mu(1-\delta)\\
&\beta= \varepsilon -e Q, \quad \mu=r_+m, \quad \varepsilon=r_+E, \quad \delta=\frac{r_-}{r_+}
\end{split}
\end{equation}

By introducing the functions $\hat f^\pm_{E,\kappa}(x)$ defined by \cite{sporea2,cota1}
\begin{equation}\label{sol8}
\begin{split}
& \hat f^+_{E,\kappa} = \frac{1}{2}\frac{f^+_{E,\kappa}}{\sqrt{\epsilon+\mu}} +\frac{1}{2}\frac{f^-_{E,\kappa}}{\sqrt{\mu-\epsilon}} \\
& \hat f^-_{E,\kappa} = -\frac{1}{2}\frac{f^+_{E,\kappa}}{\sqrt{\epsilon+\mu}} +\frac{1}{2}\frac{f^-_{E,\kappa}}{\sqrt{\mu-\epsilon}}
\end{split}
\end{equation}
one can solve analytically the resulted equations for the new radial wave functions $\hat f^\pm(x)$.

A bound/quasibound state is in fact a normalisable (or at least square integrable) solution to the massive Dirac equation that behaves as an ingoing wave at the horizon, while in the far field region falls off exponentially. Thus, these states are different from the quasinormal modes that behave asymptotically as purely outgoing waves. As a result the quasiresonances obtained when studying the spectrum of quasinormal modes (obtained for example in Refs. \cite{QNM17, QNM17a, QNM18}) are not related with the quasibound states computed here. The discrete quantum modes that give rise to this kind of states with discrete energy levels are obtained by solving the above radial Dirac equation in the case $\mu>\varepsilon$. As in Ref. \cite{cota1} where the Schwarzschild discrete modes were found, by applying the same method and after some computations the following particular solutions of the Dirac equation in a Reissner-Nordstr\"om background are found \cite{sporeathesis}
\begin{equation}\label{bond1}
\begin{split}
&\hat f^+_b=C^+(2\nu)^{s+\frac{1}{2}}x^{2s}e^{-\nu x^2}\, _1F_1(s-p+1, 2s+1, 2\nu x^2) \\
&\hat f^-_b=C^-(2\nu)^{s+\frac{1}{2}}x^{2s}e^{-\nu x^2}\, _1F_1(s-p, 2s+1, 2\nu x^2)
\end{split}
\end{equation}
that depend on the parameters
\begin{equation}\label{bond2}
s=\sqrt{\kappa^2+\zeta^2-\beta^2}, \qquad p=\frac{\beta\varepsilon-\zeta\mu}{\nu}, \qquad \nu=\sqrt{\mu^2-\varepsilon^2}
\end{equation}
while the two normalisation constants must satisfy
\begin{equation}\label{bond3}
\frac{C^-}{C^+}=\frac{\nu(s+p)}{\kappa\nu+\beta\mu-\zeta\varepsilon}
\end{equation}

In the next section we will use these solutions to find bound/quasibound states in the geometry of a Reissner-Nordstr\"om black hole.

\section{Energy of the quasibound states}\label{sec3}

The solutions (\ref{bond1}) obtained in the previous section are similar to the wave functions for a Dirac particle moving in a Coulomb potential \cite{Greiner}. Moreover, in order for the solutions to be square integrable the hypergeometric functions must reduce to simple polynomials.  Using this observation, the bound state energies of the fermions in the geometry of a Reissner-Nordstr\"om black hole can be found by imposing the standard quantization condition that the first argument of the hypergeometric function be a negative integer
\begin{equation}\label{bond4}
s-p=-n_r
\end{equation}
with $n_r=0,1,2,3...$ the radial quantum number, related to the principal quantum number by $n_r=n-|\kappa|=n-j-1/2$. Substituting now the parameters defined in eq. (\ref{bond2}), the quantization condition (\ref{bond4}) becomes
\begin{equation}\label{bond5}
\sqrt{\kappa^2+\zeta^2-\beta^2}-\frac{\beta\varepsilon-\zeta\mu}{\nu}=-n_r
\end{equation}
This is a messy equation for the energy $E$ (contained in $\varepsilon$) of the bound states. The existence of a nonvanishing inward current at the black hole horizon \cite{dolan} implies that the bound state must decay and it's more appropriate to name these states quasibound states. For that reason the energy of the quasibound state will have a real and an imaginary part \cite{dolan,huang}. However, eq. (\ref{bond5}) can have also pure real solutions for particular sets of parameters, but we must keep in mind that eq. (\ref{bond5}) is only approximative. In Section \ref{sec4} we will focus on finding and discussing only the real solutions to eq. (\ref{bond5}).

Let us now briefly discuss analytically the limiting cases of eq. (\ref{bond5}). After some manipulations, relation (\ref{bond5}) can be brought into the form \cite{sporeathesis}
\begin{equation}\label{bond6}
\frac{\varepsilon}{\mu}=\left[ 1-\left( \frac{\beta\frac{\varepsilon}{\mu}-\zeta}{n_r+\sqrt{\kappa^2+\zeta^2-\beta^2}} \right)^2 \right]^{\frac{1}{2}}
\end{equation}

Furthermore, if we assume that the energy of the quasibound state is close to the rest energy of the fermion $mc^2$, then by taking the limit $\varepsilon\to\mu$ in the right hand side part of eq. (\ref{bond6}), the following approximative expression for the energy of the quasibound state of a fermion in Reissner-Nordstr\"om geometry is obtained
\begin{equation}\label{bond7}
\frac{E}{mc^2}=\left[ 1-\left( \frac{\mu-eQ-\zeta}{n_r+\sqrt{\kappa^2+\zeta^2-(\mu-eQ)^2}} \right)^2 \right]^{\frac{1}{2}}
\end{equation}

The Schwarzschild result for the energy of a quasibound state, first derived in ref. \cite{cota1}, is recovered by canceling the black hole's electric charge i.e. making $Q=0$ in eq. (\ref{bond7}) to obtain
\begin{equation}\label{bond8}
\frac{E}{mc^2}=\left[ 1-\frac{\mu^2}{4\left(n_r+\sqrt{\kappa^2-\frac{3}{4}\mu^2}\,\right)^2} \right]^{\frac{1}{2}}
\end{equation}

Eq. (\ref{bond7}) reduces in the limit $M\to 0$ to the formula for the discrete energy levels of the relativistic Dirac-Coulomb problem (see ref. \cite{Landau4})
\begin{equation}\label{bond8a}
\openup 2\jot
\begin{split}
\frac{E}{mc^2}&=\left[ 1-\left( \frac{eQ}{n_r+\sqrt{\kappa^2-(eQ)^2}} \right)^2 \right]^{\frac{1}{2}} \\
&\approx \left[ 1+\frac{Z^2\alpha^2}{\left(n_r+\sqrt{\kappa^2-Z^2\alpha^2}\right)^2} \right]^{-\frac{1}{2}}, \qquad Q=Ze=Z\sqrt{\alpha}
\end{split}
\end{equation}
Moreover, if $M=0$ the approximative solutions (\ref{bond1}) also reduce to the exact solutions of the Dirac-Coulomb problem \cite{Landau4}.

\section{Discussion of the results}\label{sec4}

\begin{figure}[h!t]
	\centering
	\includegraphics[scale=0.415]{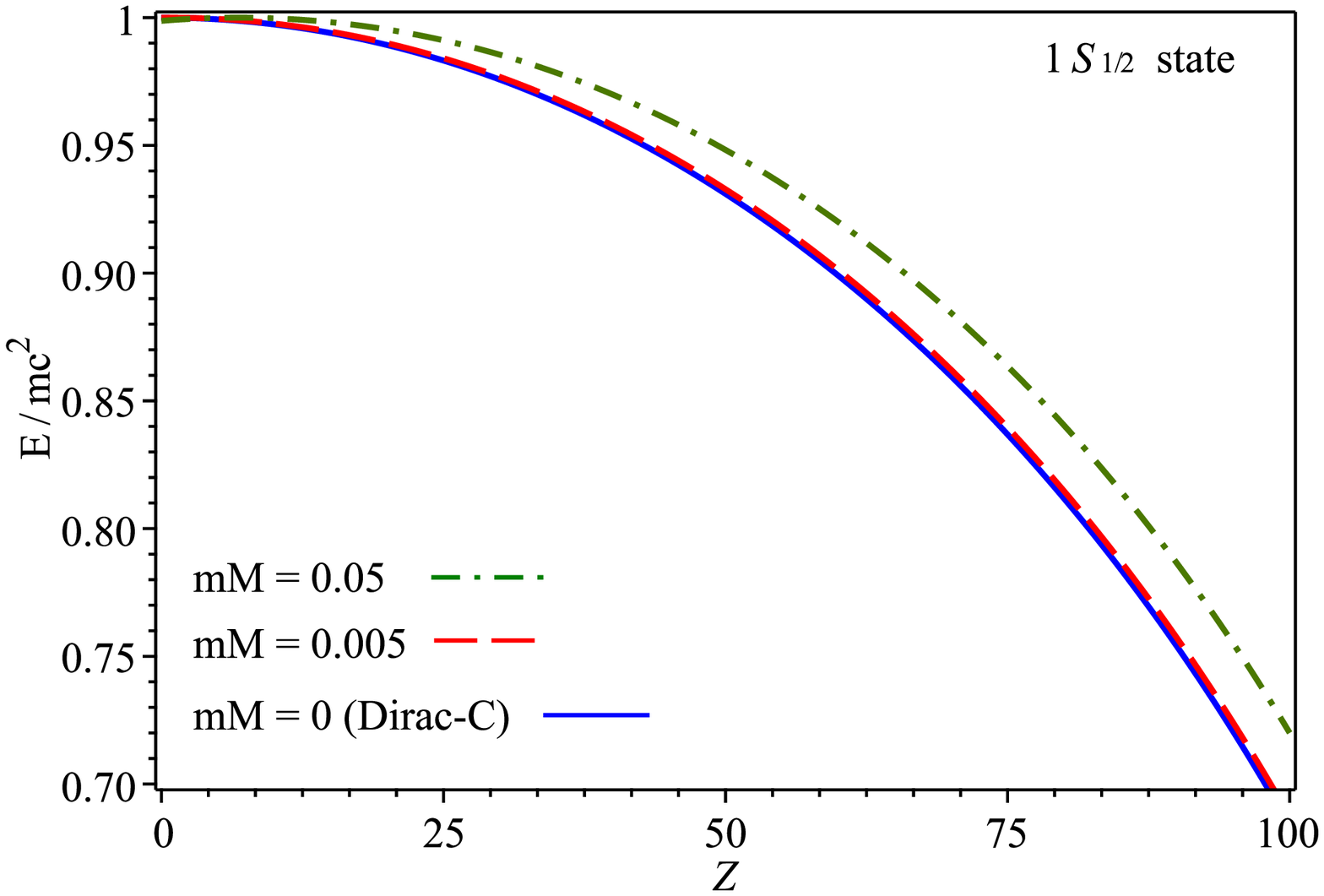}
	\includegraphics[scale=0.415]{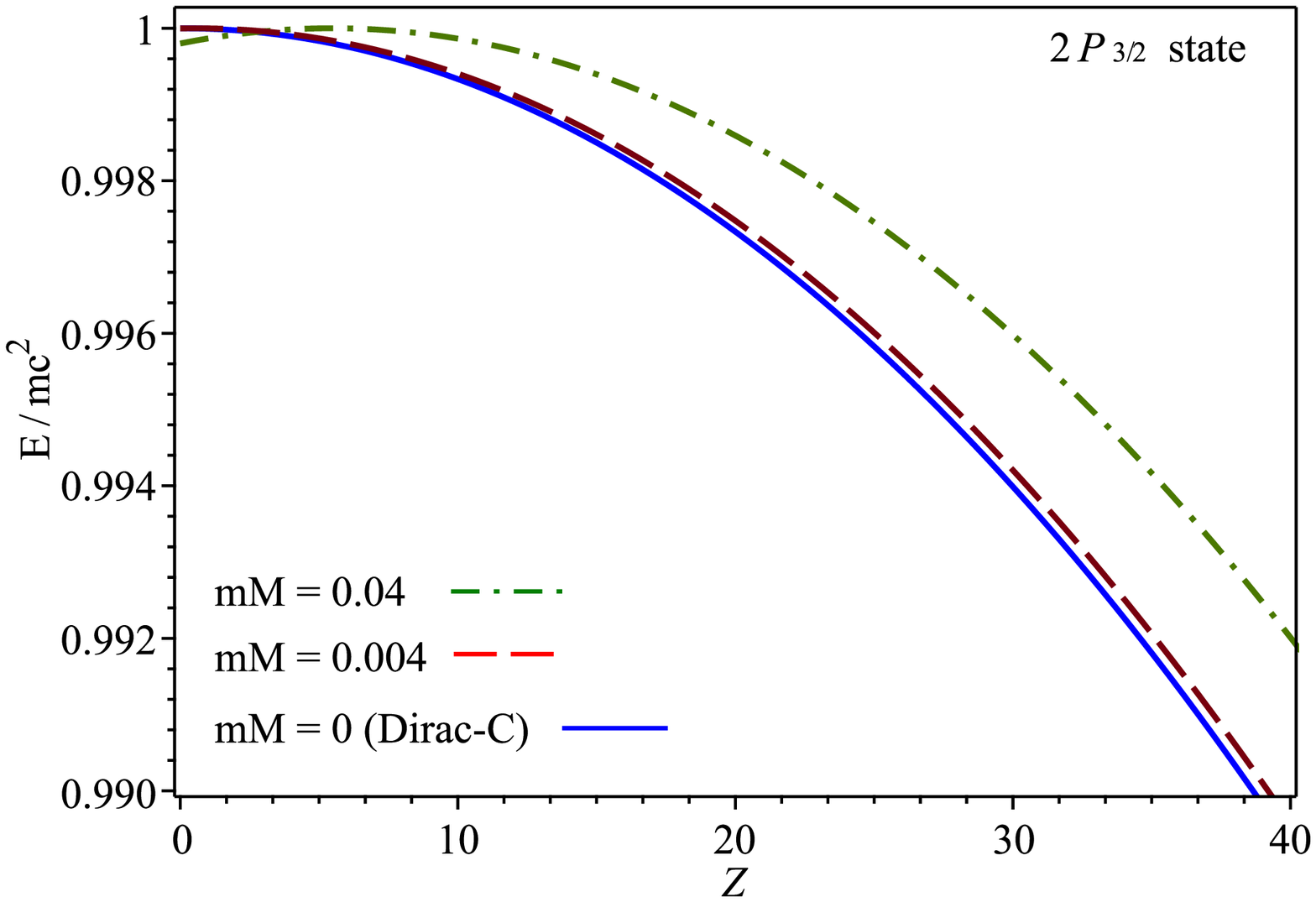}
		\caption{Comparison of the Reissner-Nordstr\"om ground state energy with the relativistic Dirac-Coulomb energy for the $1S_{1/2}$ state (left panel), respectively for the $2P_{3/2}$ state (right panel). }
	\label{fig.1}
\end{figure}

\begin{figure}[h!t]
\centering
\includegraphics[scale=0.415]{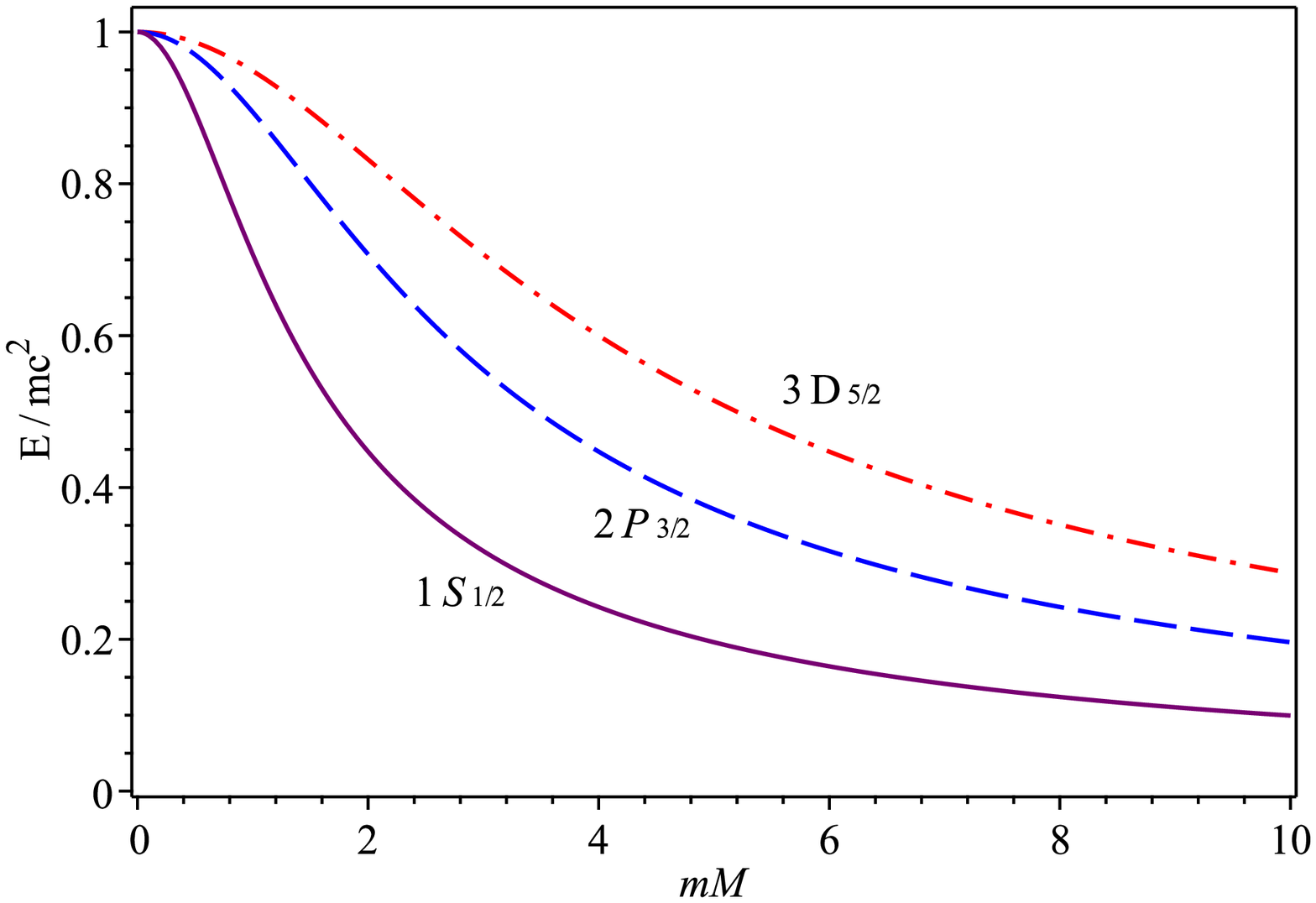}
\includegraphics[scale=0.415]{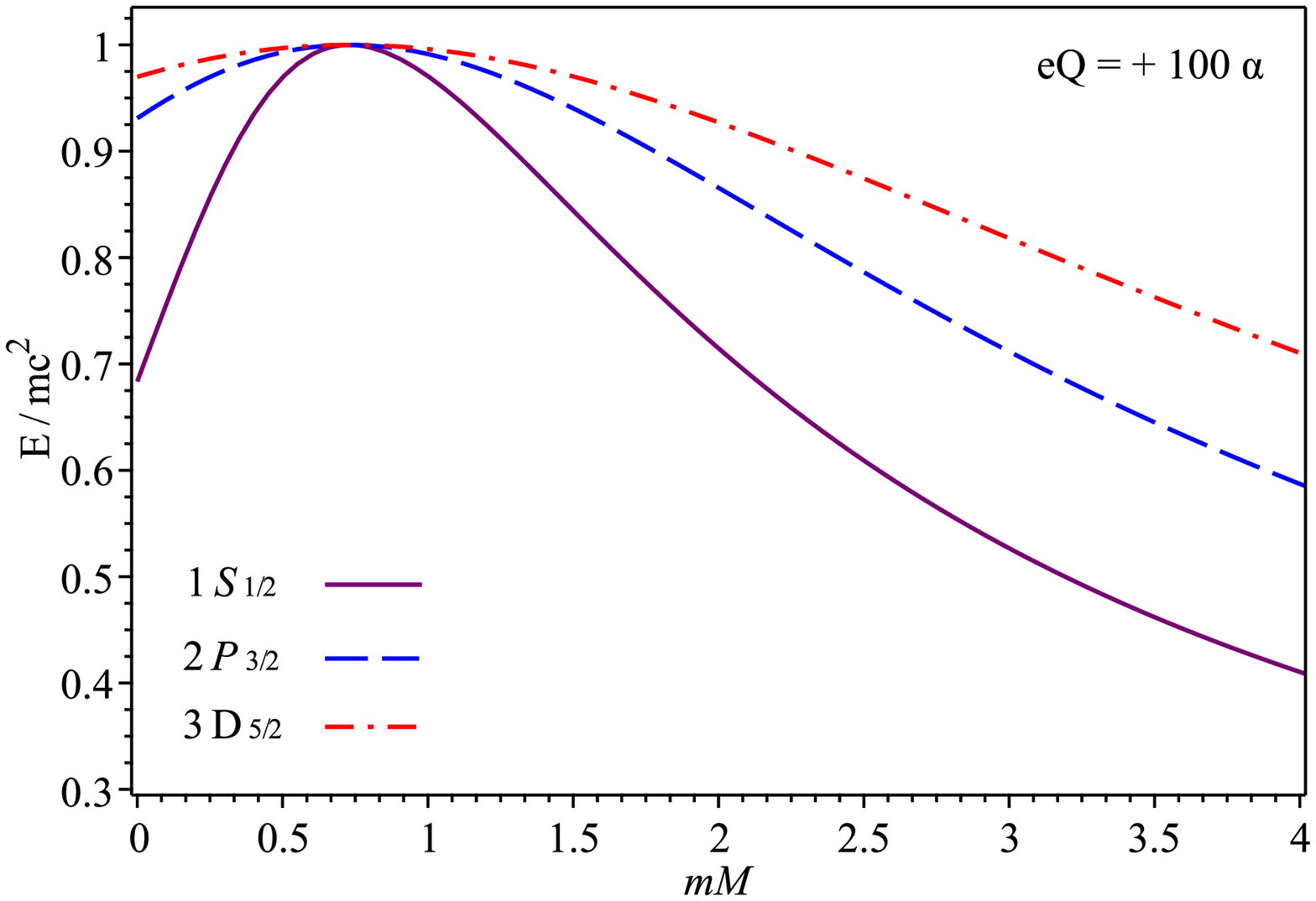}
\caption{The energy spectra for Schwarzschild quasibound states (left panel), respectively for Reissner-Nordstr\"om quasibound states (right panel) as functions of the gravitational coupling $\alpha_g=\frac{mMG}{\hbar c}$ for states with $n_r=0$.}
\label{fig.2}
\end{figure}

\begin{figure}[h!t]
	\centering
	\includegraphics[scale=0.415]{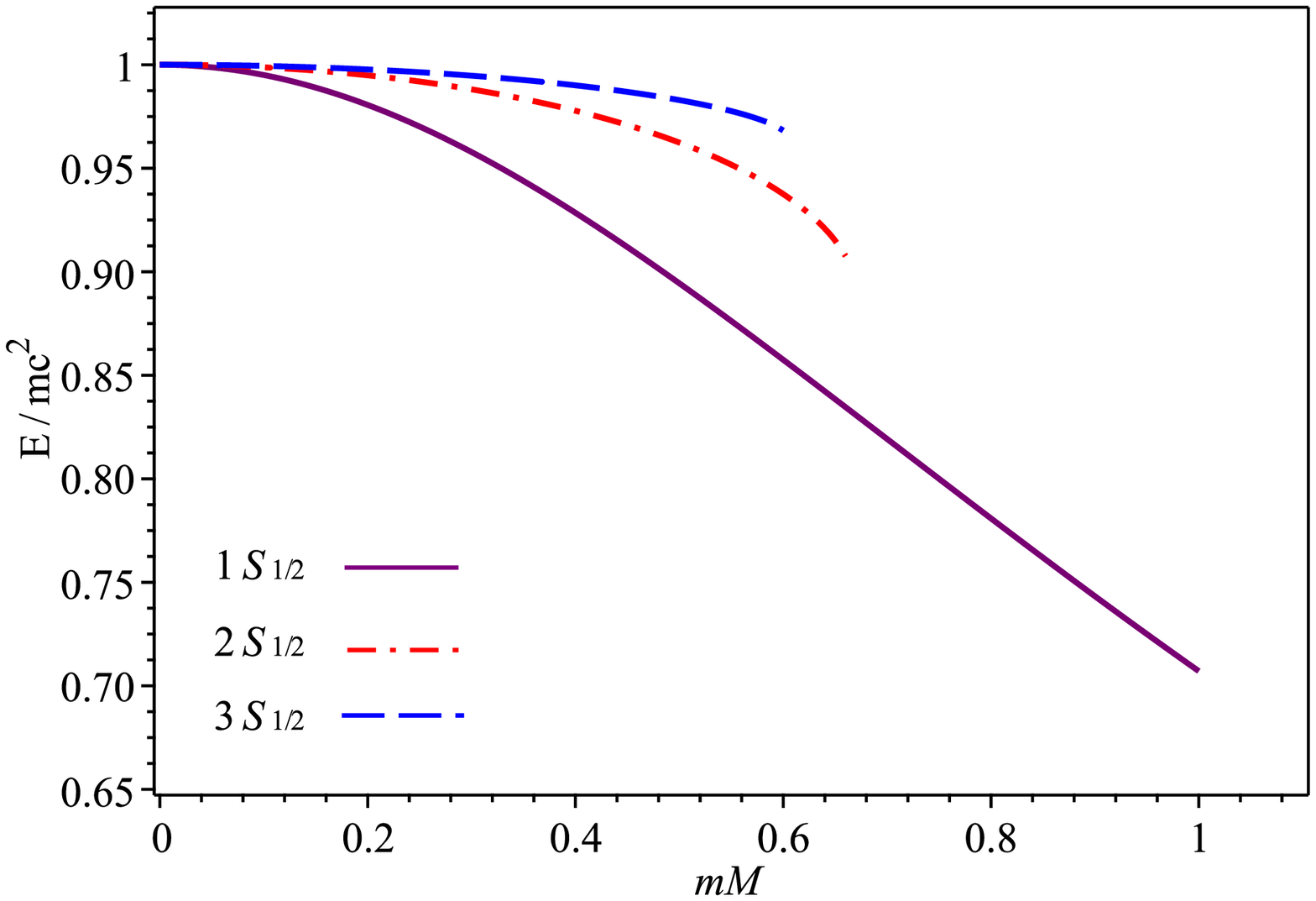}
	\includegraphics[scale=0.415]{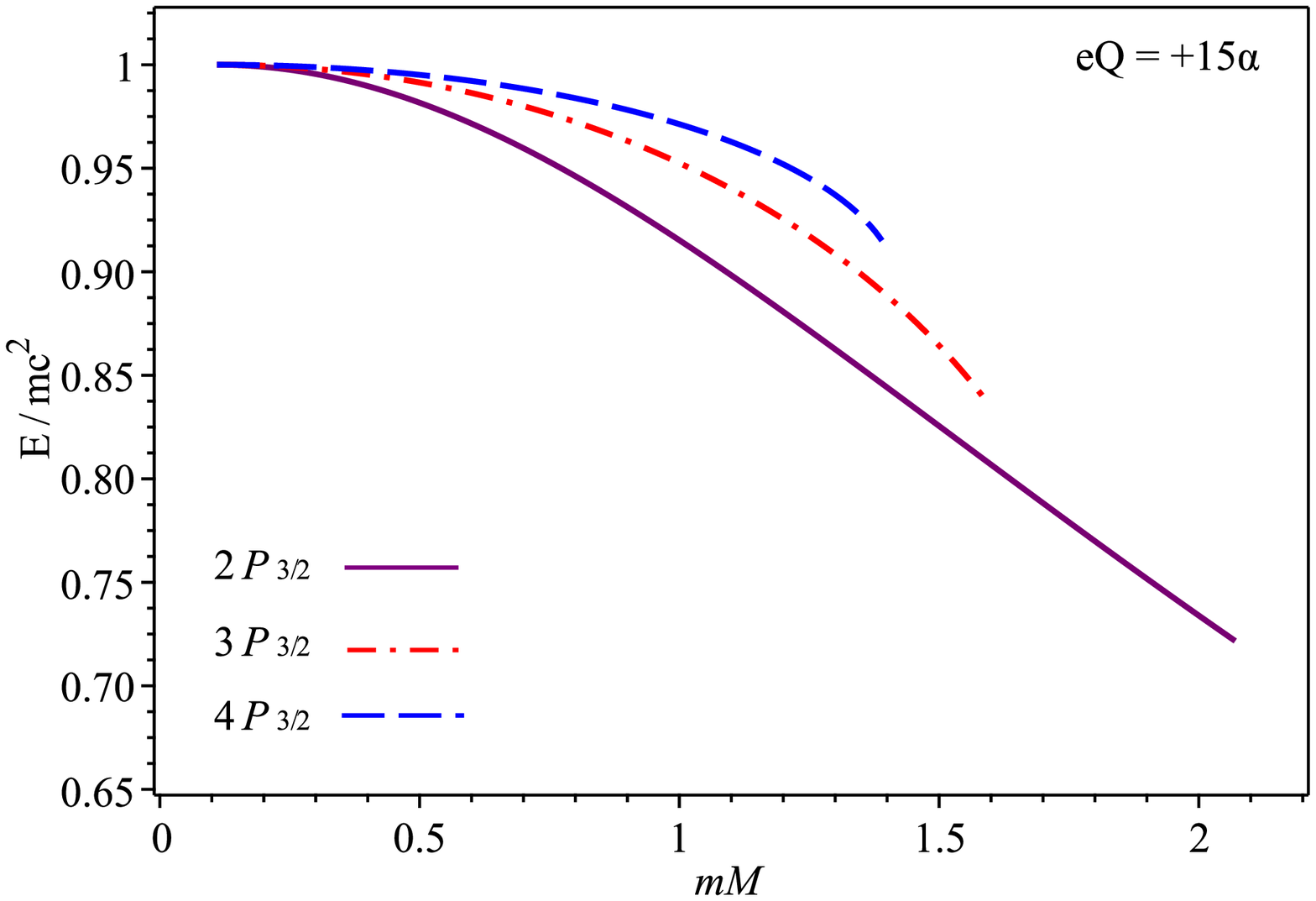}
	\caption{The energy spectra for Schwarzschild ({\it left panel}) and Reissner-Nordstr\"om ({\it right panel}) quasibound states as a function of the gravitational coupling $\alpha_g=\frac{mMG}{\hbar c}$ for states with $n_r\neq0$.}
	\label{fig.3}
\end{figure}

In Figs. \ref{fig.1}-\ref{fig.5} we present the real solutions of the energy $E/mc^2$ as a function of the gravitational coupling $\alpha_g=\frac{mMG}{\hbar c}$ (labeled on the plots as $mM$) for several quasibound states. The plots are obtained by solving analytically (for $n_r=0$) or numerically (if $n_r\neq0$) equation (\ref{bond5}) and the states are labeled with the standard spectroscopic notation $nL_j$. Thus the set of quantum numbers $(n=1, l=0, j=1/2, \kappa=-1)$ corresponds to the energy state labeled $1S_{1/2}$; the set $(n=2, l=1, j=1/2, \kappa=1)$ is attributed to the state $2P_{1/2}$; $(n=3, l=2, j=3/2, \kappa=2)$ to the state $3D_{3/2}$ and so on.
Because eq. (\ref{bond5}) depends only on $\kappa^2$ the states of the form $nS_{1/2}$ and $nP_{1/2}$ will give (in the approximation used here) states with the same energy i.e. we will have degenerate levels.


Let us start by analysing what happens with the energy of the ground state, that has $n_r=0$. In this case eq. (\ref{bond5}) has two solutions given by the following expression
\begin{equation}\label{E01}
\frac{E_0}{mc^2}=\frac{mM\cdot eQ\pm\sqrt{\kappa^2\left[ (mM)^2+\kappa^2-(eQ)^2 \right]}}{(mM)^2+\kappa^2}
\end{equation}
By imposing the condition
\begin{equation}
-\sqrt{\kappa^2+(mM)^2} < eQ < \sqrt{\kappa^2+(mM)^2}
\end{equation}
the ground state will always have a real energy, otherwise the energy of the state becomes complex. For certain values of the parameters in eq. (\ref{E01}) and after imposing the condition $0<\frac{E_0}{mc^2}<1$, only one of the two solutions is physical.

In Fig. \ref{fig.1} we compare the energy of the Reissner-Nordstr\"om ground state with the relativistic Dirac-Coulomb energy for the $1S_{1/2}$ and $2P_{3/2}$ stats. We observe that the two energies agree quite well for a small value of the gravitational coupling $mM$. However, as the mass of the black hole is increased, the energy of the ground states start to depart from each other.

Fig. \ref{fig.2} shows the energy of several quasibound states as a function of the gravitational coupling $\alpha_g\equiv mM$ (in units of $c=1,\,G=1,\,\hbar=1$) for the Schwarzschild $1S_{1/2}$, $2P_{3/2}$, $3D_{5/2}$ states in the left panel, while in the right panel we present the same states but for a Reissner-Nordstr\"om black hole with $eQ=100\alpha$. In each case, for a fixed value of the parametar $mM$ the energy of the state is higher as the principal quantum number $n$ is increased. Moreover, in the case of Reissner-Nordstr\"om quasibound states a maximum appears, who's width increases for higher states. The maximum is present only for the case $eQ>0$ (see also Fig. \ref{fig.5}).

For states with $n_r\neq 0$, equation (\ref{bond5}) can not be solved analytically. Searching for numerical solution using Maple, we have found that for values of $mM\leq \alpha_0$, eq. ({\ref{bond5}) has only one positive real solution. For values outside this interval, we were unable to find numerical solutions with Maple. We found that for the states with $n>|\kappa|$ the value of $\alpha_0<n$. In Fig. \ref{fig.3} we present the energy of the $1S_{1/2}$, $2S_{1/2}$, $3S_{1/2}$ Schwarzschild quasibound states (in the left panel), respectively the energy of the $2P_{3/2}$, $3P_{3/2}$, $4P_{3/2}$ Reissner-Nordstr\"om quasibound states (in the right panel).

\begin{figure}[h!t]
	\centering
	\includegraphics[scale=0.415]{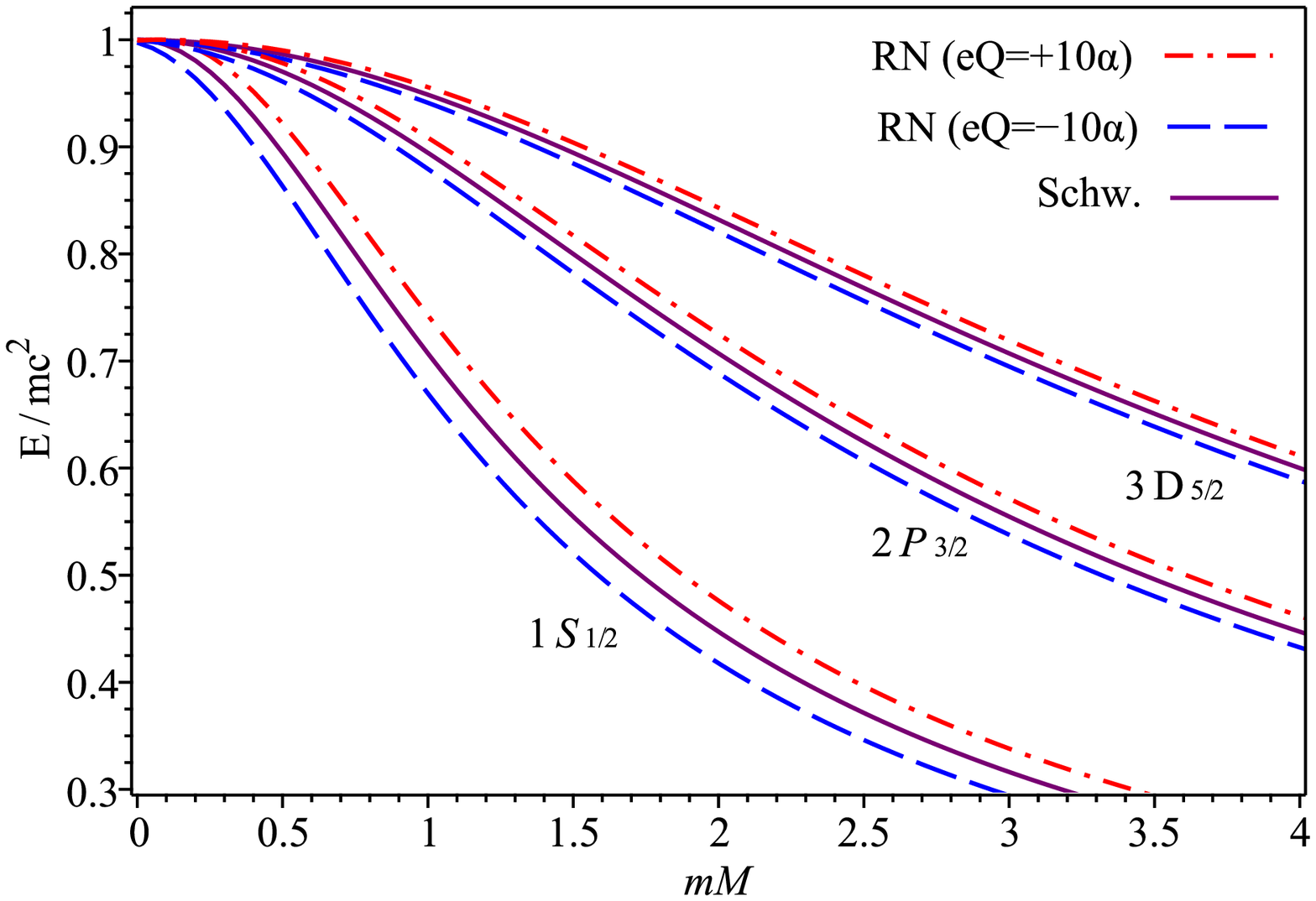}
	\includegraphics[scale=0.415]{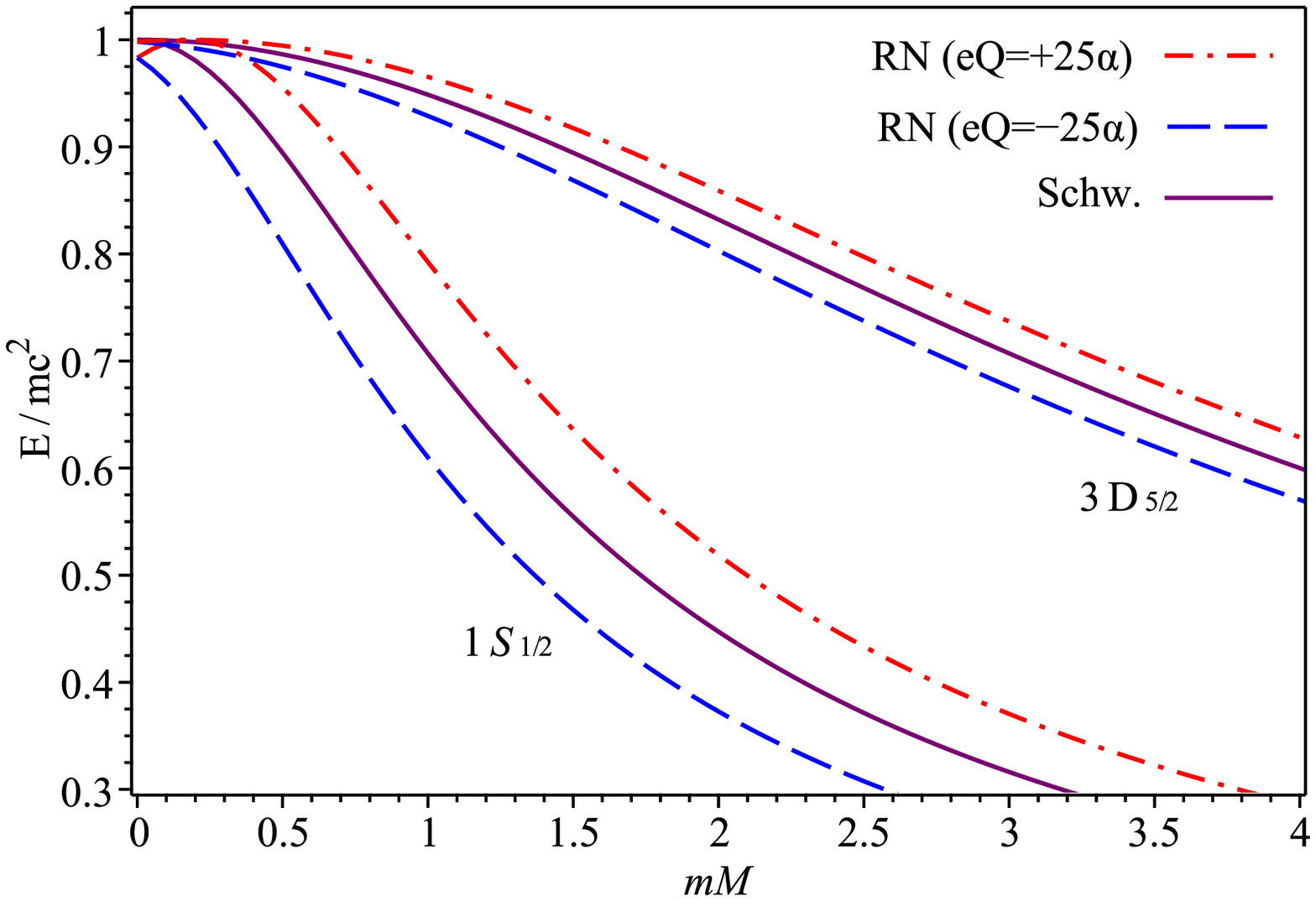}
	\caption{Comparison between the real part of the energy spectra of a Schwarzschild and Reissner-Nordstr\"om $1S_{1/2}$, $2P_{3/2}$, $3D_{5/2}$ quasibound states for: $eQ=\pm 10\alpha$ ({\it left panel}), respectively for $eQ=\pm25\alpha$ ({\it right panel}). We observe that the spectra of RN quasibound states compared with the Schwarzschild one, is higher if the fermion and the black hole have the same type of charge (i.e. $eQ>0$) ({\it left panel}), respectively is lower for the opposite case for which $eQ<0$ ({\it right panel}). }
	\label{fig.4}
\end{figure}

\begin{figure}[h!t]
\centering
\includegraphics[scale=0.415]{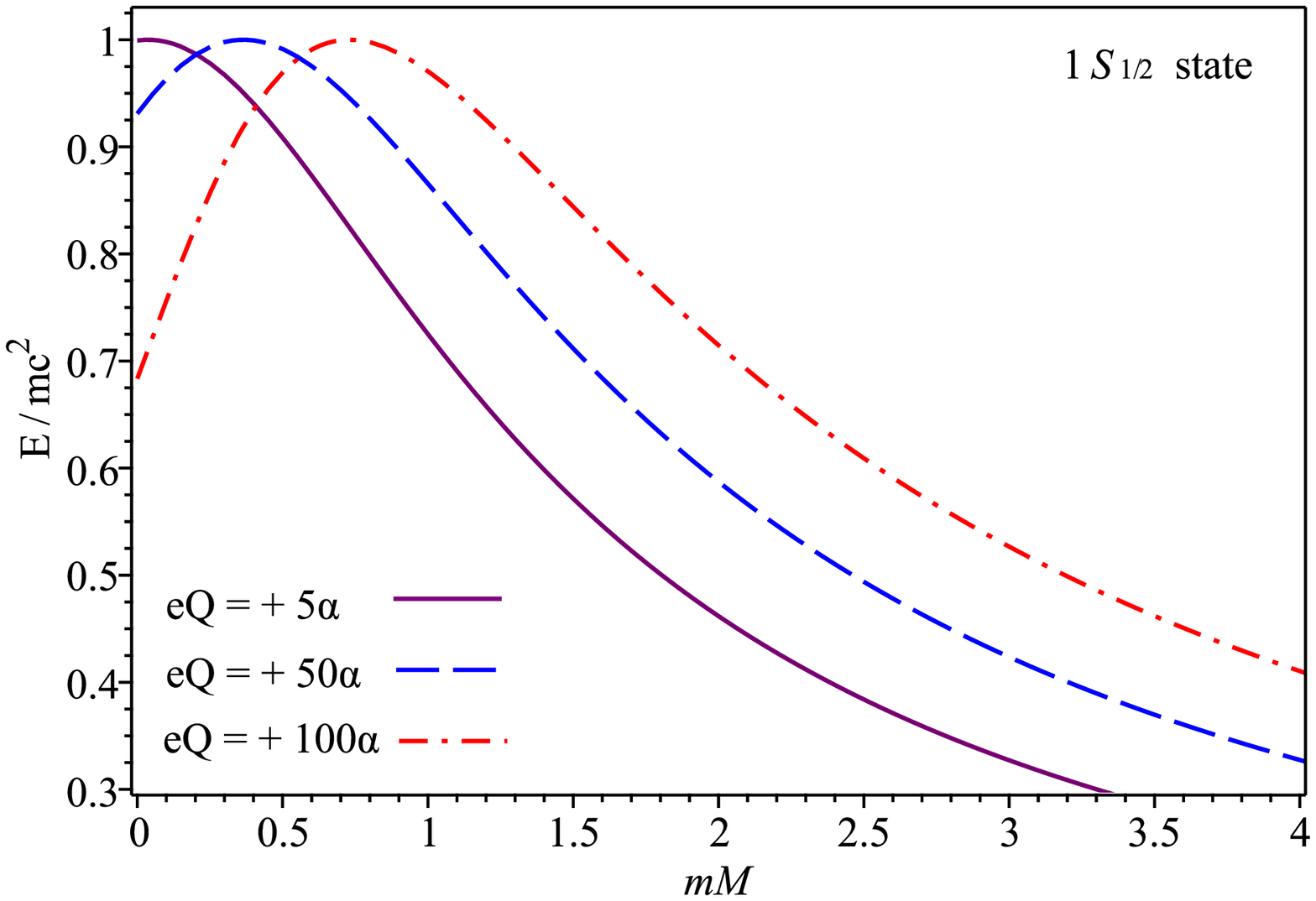}
\includegraphics[scale=0.415]{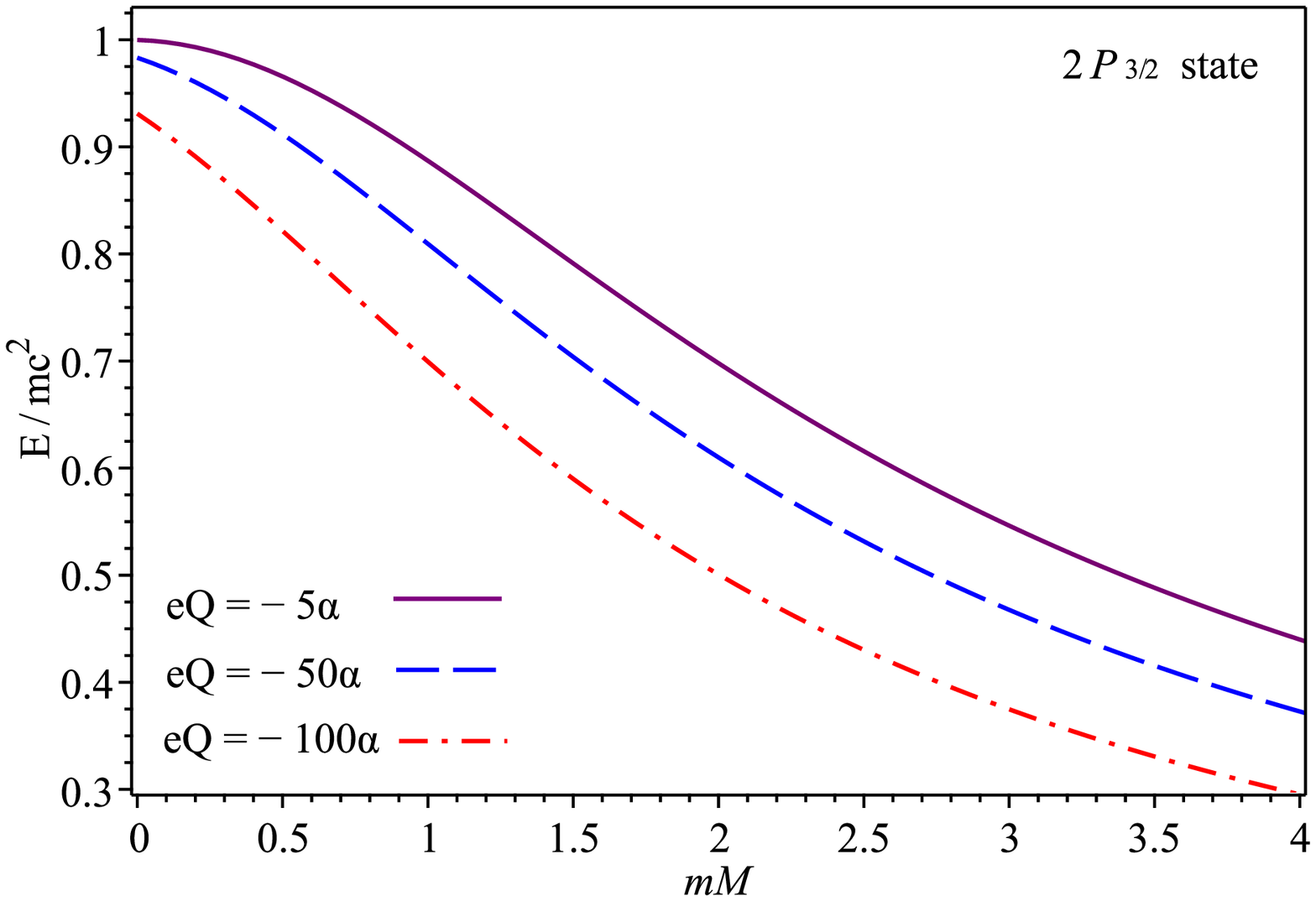}
\caption{The energy spectra of the $1S_{1/2}$ quasibound  state for a Reissner-Nordstr\"om black hole charged with positive or negative charges. In the left panel we observe that as $eQ$ increases the spectra can have the same energy at two different values of the gravitational coupling $\alpha_g=mMG/\hbar c$.}
\label{fig.5}
\end{figure}

In Fig. \ref{fig.4}  we compare the energies of Schwarzschild and Reissner-Nordstr\"om quasibound states for $1S_{1/2}$, $2P_{3/2}$ and $3D_{5/2}$ states. We observe that for a fixed value of the gravitational coupling (i.e. the parameter $mM$) the effect of adding only a few negative charges to the black hole ($eQ>0$), assuming a negatively charged fermion, will result in a Reissner-Nordstr\"om quasibound state with a higher energy compared with the energy of the same Schwarzschild quasibound state. Furthermore, adding positive charges ($eQ<0$) to the black hole will lower the energy of the Reissner-Nordstr\"om quasibound state. If more charge is added to the black hole, the difference in energy will increase further, as can be observed in the right panel of Fig. \ref{fig.4}.

Analyzing Fig. \ref{fig.5} an interesting effect can be observed (see the left panel), namely that a given quasibound state (the $1S_{1/2}$ state in our example) can have the same energy for two different values of the gravitational coupling $\alpha_g$. This effect can be observed only for black holes that have the same type of charge as the fermion (i.e. $eQ>0$). This is the case corresponding to electromagnetic repulsion, when the fermion scattering intensity takes higher values \cite{sporea2}. Furthermore, as one varies the parameter $mM$ the energy of a RN quasibound state increases for small values of $mM$ up to a maximum and then it starts again to decrease. In the case of $eQ<0$ these effects are not observed as can be seen from the figure in the right panel.

\begin{figure}[h!t]
\centering
\includegraphics[scale=0.45]{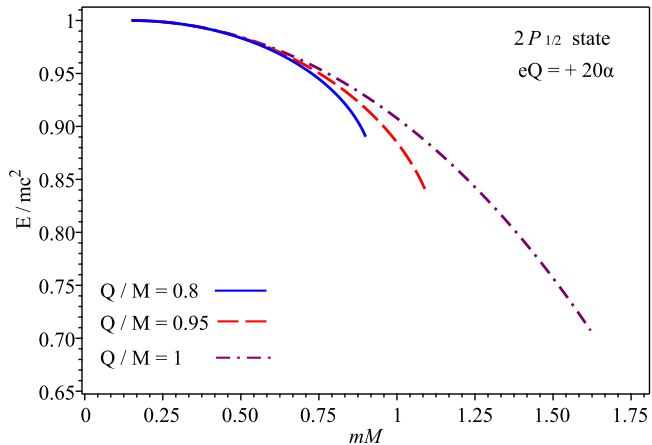}
\caption{The energy spectra of the $2P_{1/2}$ quasibound  state in the Reissner-Nordstr\"om extremal limit $M=Q$.}
\label{fig.6}
\end{figure}

In the last Figure \ref{fig.6} we present the energy of the $2P_{1/2}$ quasibound state as a function of the gravitational coupling $mM$ for high values of the ratio $Q/M$ between the black hole charge and it's mass, including also the extremal limit $Q=M$. We observe that for small values of $mM$ the quasibound states are almost degenerate and as the value of $mM$ is increased the quasibound state with the highest energy is the one corresponding to the extremal Reissner-Nordsrt\"om black hole.

\section{Conclusions}\label{concl}

In summary, we have studied the (quasi)bound states of the Dirac filed in Schwarzschild and Reissner-Nordstr\"om black hole geometries, with a focus on the later one. These states were obtained by applying a quantization condition to the discrete quantum modes of the Dirac equation in the two geometries. For the ground state (having the radial quantum number $n_r=0$) we were able to find an analytical expression for the energy of the state. For the states with $n_r\neq0$ we used Maple to compute numerically the energy of the states. We have found that the Reissner-Nordstr\"om quasibound states have higher energies compared with the Schwarzschild quasibound states if the black hole and the fermion have chargees with the same sign, otherwise the energy of the state is lower.

\begin{acknowledgements}
This work was supported by a grant of Ministery of Research and Innovation, CNCS - UEFISCDI, project number PN-III-P1-1.1-PD-2016-0842, within PNCDI III. I am grateful to Professor I.I. Cotaescu for many discussions on this topic and for suggestions that helped improve the final version of the manuscript. I would also like to thank C. Crucean for reading the manuscript.

\end{acknowledgements}


\begin{thebibliography}{90}

\bibitem{futterman}
J.A.H., Futterman, F.A., Handler,  R.A.,  Matzner,  Scattering from black holes (Cambridge University Press, Cambridge, 1988).

\bibitem{dolan}
C. J. L. Doran, A. N. Lasenby, S. Dolan, and I. Hinder, Phys. Rev. D 71, 124020 (2005).

\bibitem{dolan11}
S. Dolan, C. Doran and A. Lasenby, Phys. Rev. D {\bf 74}, 064005 (2006).

\bibitem{brito}
M.A. Anacleto, F.A. Brito, S.J.S. Ferreira, E. Passos, Phys. Lett. B 788 (2019) 231-237.

\bibitem{unruh}
W. G. Unruh, Phys. Rev. D14, 3251 (1976).

\bibitem{Rogatko}
M. Rogatko and A. Szyplowska, Phys. Rev. D 79, 104005 (2009).

\bibitem{Gaina1}
A.B. Gaina, Moscow VINITI, No. 1970-80 Dep., 20 pp. (1980).

\bibitem{Matzner1}
P. Anninos, C. DeWitt-Morette, R. A. Matzner, P. Yioutas, and T. R. Zhang, Phys. Rev. D 46, 4477 (1992).

\bibitem{Matzner2}
R. A. Matzner, C. DeWitte-Morette, B. Nelson, and T.-R. Zhang, Phys. Rev. D31, 1869 (1985).

\bibitem{Ford1}
K. W. Ford and J. A. Wheeler, Ann. Phys. (N.Y.) 7, 259 (1959).

\bibitem{Ford2}
K. W. Ford and J. A. Wheeler, Ann. Phys. (N.Y.) 7, 287 (1959).

\bibitem{konoplya}
R.A. Konoplya, A. Zhidenko, Phys. Rev. D 76, 084018 (2007).

\bibitem{sanchez}
N.G. Sanchez, J. Math. Phys. (N.Y.) 17, 688 (1976)

\bibitem{sanchez1}
N.G. Sanchez, Phys. Rev. D 18, 1798 (1978)

\bibitem{anderson}
N. Andersson, Phys. Rev. D 52, 1808 (1995).

\bibitem{batic}
D. Batic, N.G. Kelkar, M. Nowakowski, Eur. Phys. J. C 71, 1831 (2011).

\bibitem{crispino11}
C. F. B. Macedo, E. S. de Oliveira, L. C. B. Crispino, Phys.Rev. D 92 (2015) no.2, 024012.

\bibitem{crispino12}	
L.C.B. Crispino, A. Higuchi, E. S. Oliveira, Phys. Rev. D 80 (2009) 104026.

\bibitem{oliverira}
E. S. de Oliveira, Eur. Phys. J. C (2018) 78:876.


\bibitem{sporea1}
I.I. Cot\u aescu, C. Crucean, C.A. Sporea, Eur. Phys. J. C {\bf 76}:102 (2016).

\bibitem{sporea2}
I.I. Cot\u aescu, C. Crucean, C.A. Sporea, Eur. Phys. J. C {\bf 76}:413 (2016).

\bibitem{sporea3} C.A. Sporea, Chinese Physics C, Vol. 43, No. 3 (2019) 035104.

\bibitem{wu-dai1}
Wen-Du Li, Yu-Zhu Chen, Wu-Sheng Dai, Phys. Lett. B 786 (2018) 300-304.

\bibitem{wu-dai2}
Yuan-Yuan Liu, Wen-Du Li, Wu-Sheng Dai, arXiv:1902.01054.



\bibitem{QNM1}
H.-J. Blome and B. Mashhoon, Phys. Lett. A100, 231 (1984).

\bibitem{QNM2}
B.F.  Schutz  and  C.M.  Will,  Astrophys.  J.  Lett. 291,  L33 (1985).

\bibitem{QNM3}
S. Iyer and C.M. Will, Phys. Rev. D35, 3621 (1987).

\bibitem{QNM4}
H. T. Cho, Phys. Rev. D 68, 024003 (2003).

\bibitem{QNM5}
R.A. Konoplya, Phys. Rev. D66, 084007 (2002).

\bibitem{QNM6}
L.E. Simone,  C.M. Will, Classical and Quantum Gravity, Vol. 9, (1992) 012, pp. 963-977.

\bibitem{QNM7}
V. Cardoso, R. Konoplya, and J. P. S. Lemos, Phys. Rev. D 68 044024 (2003).



\bibitem{QNM8}
J. Jing, Phys. Rev. D 69 084009 (2004).


\bibitem{QNM9}
J. Jing and Q. Pan, Nuclear Physics B 728, 109 (2005).

\bibitem{QNM10}
E. Berti, V. Cardoso, and A. O Starinets, Class. Quantum Grav. 26, 163001 (2009).


\bibitem{QNM11}
N. Varghese and V. C. Kuriakose, Gen. Relativ. Gravit. 41, 1249 (2009).

\bibitem{QNM12}
K. D. Kokkotas, R. A. Konoplya, and A. Zhidenko, Phys. Rev. D 83 024031 (2011).

\bibitem{QNM13}
A. Flachi and J. P. S. Lemos, Phys. Rev. D 87. 024034 (2013).

\bibitem{QNM14}
M. Richartz and D. Giugno, Phys. Rev. D 90, 124011 (2014).

\bibitem{QNM15}
Chen Wu, Int. J. Mod. Phys. D 26, 1750111 (2017).

\bibitem{QNM16}
G. Panotopoulos and Á. Rincón, Int. J. Mod. Phys. D 27, 1850034 (2018).

\bibitem{QNM17}
J. L. Blázquez-Salcedo and C. Knoll, Phys. Rev. D 97, 044020 (2018).


\bibitem{QNM17a}
J. L. Blázquez-Salcedo and C. Knoll, Class. Quantum Grav. 36 105012 (2019).


\bibitem{QNM18}
R. A. Konoplya and A. Zhidenko, Phys. Rev. D 97, 084034 (2018).

\bibitem{QNM19}
B. Gwak, arXiv:1812.04923 [gr-qc].


\bibitem{QNM20}
C. F. B. Macedo, V. Cardoso, L. C. B. Crispino, P. Pani, Phys.Rev. D 93 (2016) no.6, 064053.


\bibitem{barranco}
J. Barranco, A. Bernal, J. C. Degollado, A. Diez-Tejedor,M.
Megevand, M. Alcubierre, D. Núñez, and O. Sarbach, Phys. Rev. Lett. 109, 081102 (2012).


\bibitem{herdeiro}
J. C. Degollado and C. A. R. Herdeiro, Gen. Relativ. Gravit. 45, 2483 (2013).

\bibitem{xiang-nan}
Xiang-Nan Zhou, Xiao-Long Du, Ke Yang, and Yu-Xiao Liu, Phys. Rev. D 89, 043006 (2014)

\bibitem{rez1}
 Jing Ji-Liang and Pan Qi-Yuan 2008 {\it Chinese Phys. B} 17 1985

\bibitem{rez2}
J. Barranco, A. Bernal, J. C. Degollado, A. Diez-Tejedor,M. Megevand, M. Alcubierre, D. Núñez, and O. Sarbach, Phys. Rev. D 84, 083008 (2011).

\bibitem{rez3}
H.S. Vieira, V.B. Bezerra, Annals of Physics, Vol. 373, October 2016, pp. 28-42.

\bibitem{rez4}
J. Jing, Q. Pan, and X. He, Int. J. Mod. Phys. D 16 81 (2007).

\bibitem{rez5}
S. Hod, Eur. Phys. J. C (2017) 77:351

\bibitem{rez6}
Y. Decanini, A. Folacci and B. Raffaelli, Phys. Rev. D 84, 084035 (2011).

\bibitem{dolan1}
A. Lasenby, C. Doran, J. Pritchard, A. Caceres, and S. Dolan, Phys. Rev. D 72, 105014 (2005).

\bibitem{gaina1}
A. B. Gaina and F. G. Kochorbe, Sov. Phys. JETP 65, 211 (1987).

\bibitem{gaina2}
I. M. Ternov and A. B. Gaina, Sov. Phys. J. 31, 157 (1988).

\bibitem{gaina3}
A. B. Gaina and O. B. Zaslavskii, Classical Quantum Gravity 9, 667 (1992).


\bibitem{cardoso}
R. Brito, V. Cardoso, and P. Pani, Phys. Rev. D 88, 023514 (2013).

\bibitem{dolan2}
J. G. Rosa and S. R. Dolan, Phys. Rev. D 85, 044043 (2012).

\bibitem{dolan3}
S. R. Dolan and D. Dempsey, Classical Quantum Gravity 32, 184001 (2015).

\bibitem{dolan4}
S. R. Dolan, Phys. Rev. D 76, 084001 (2007).

\bibitem{Ruffini}
T. Damour, N. Deruelle, and R. Ruffini, Lett. Nuovo Cimento 15, 257 (1976).


\bibitem{crispino}
C.L. Benone, L.C.B. Crispino, C.A.R. Herdeiro and E. Radu, Int. J. of Mod. Phy. D 24, (2015) 1542018.

\bibitem{crispino2a}
C.F.B. Macedo, L.C.B. Crispino, E. S. de Oliveira, Int. J. Mod. Phys. D 25 (2016) 1641008.


\bibitem{bond1a}
J. Jing, Phys. Rev. D 70, 065004 (2004).

\bibitem{bond2a}
S. Hod, Physics Letters B 749, (2015) pp. 167-171

\bibitem{bond3a}
J. Jin, Phys. Rev. D 70, 065004 (2004).

\bibitem{bond4a}
J. C. Degollado and C. A. R. Herdeiro, Phys. Rev. D 90, 065019 (2014).

\bibitem{bond5a}
H. Furuhashi, Y. Nambu, Progress of Theoretical Physics, Vol. 112, 983–995 (2004).

\bibitem{bond6a}
V, Cardoso, O J. C. Dias, J. P. S. Lemos, and S. Yoshida, Phys. Rev. D 70, 044039 (2004); Erratum Phys. Rev. D 70, 049903 (2004).

\bibitem{bond7a}
P. Pani, V. Cardoso, L. Gualtieri, E. Berti, and A. Ishibashi
Phys. Rev. D 86, 104017 (2012).

\bibitem{bond8a}
Y. Huang et al., Class. Quantum Grav. 34, 155002 (2017).

\bibitem{bond9a}
S. Hod, Physics Letters B, 2017, Vol. 770, pp 186

\bibitem{bond10a}
S. Patrick, A. Coutant, M. Richartz, and S. Weinfurtner, Phys. Rev. Lett. 121, 061101 (2018).

\bibitem{gaina}
I. M. Ternov, A. B. Gaina, and G. A. Chizhov, Sov. Phys. J. 23, 695 (1980).


\bibitem{cota1}
Cotaescu, I.I., Mod. Phys. Lett. A 22, 2493 (2007).

\bibitem{sporeathesis} C.A. Sporea, Fermion Scattering by Spherical Symmetric Black Holes, PhD thesis (2016).


\bibitem{Vilalba}
 Villalba, V. M., Mod. Phys. Lett. A 8 (1993) 2351-2364.

\bibitem{Greiner}
W. Greiner. {\it Relativistic quantum mechanics. Wave equations}. Springer, $3^{rd}$ ed., 2000.

\bibitem{huang}
Yang Huang, Dao-Jun Liu, Xiang-hua Zhai, and Xin-zhou Li
Phys. Rev. D 96, 065002 (2017).


\bibitem{Landau4}
V. B. Berestetski, E. M. Lifshitz and L. P. Pitaevski. Quantum Electrodynamics. Pergamon Press, Oxford, 1982






\end{thebibliography}
\end{document}